\documentclass[%
preprint,
amsmath,amssymb,
aps,showkeys,
]{revtex4-2}

\usepackage{graphicx}
\usepackage{dcolumn}
\usepackage{bm}

\graphicspath{{img/}} 
\usepackage{epsfig}
\usepackage[T2A]{fontenc}
\usepackage{latexsym}
\usepackage[utf8x]{inputenc}
\usepackage{xcolor}
\usepackage{todonotes}
\usepackage{mathtools}

\usepackage{todonotes}

\begin{document}
\title{On the influence of \\ electron velocity spread on Compton FEL operation}
\author{S.V. Anishchenko}
\affiliation{Research institute for nuclear problems\\ Bobruiskaya str., 11, 220030, Minsk, Belarus}
\email{sanishchenko@mail.ru}
\date{\today}
\begin{abstract}
	For the field amplitude, a nonlinear integro-differential equation is derived that describes the operation of a Compton FEL in the presence of electron velocity spread typical for modern facilities.
	Numerical solutions of the equation are in good agreement with particle simulations for the bunching factor less than 0.6, reproduce the frequency detuning spectrum near its maximum, and describe the amplification process up to saturation.	
	\end{abstract}
\maketitle

\section{Introduction}
One of the most important challenges in modern physics is the investigation of fast processes on atomic space-time scales using ultrashort pulses of electromagnetic radiation~\cite{Neutze2000,Chapman2014}. In this regard, the development and commissioning of X-ray free electron lasers (XFELs) is of great importance~\cite{Ackermann2007,LCLS2010,SACLA2011,SwissFEL2017,EuXFEL2020}. Over the past 20 years, since the first lasing in the soft X-ray range~\cite{Ackermann2007}, the progress achieved in this area has been truly impressive.

During this time, the radiation wavelength has decreased by two orders of magnitude, from ~13.7~\cite{Ackermann2007} to ~0.634~\r{A}~\cite{Tanaka2012}, the duration of X-ray pulses has stepped into the attosecond range~\cite{Huang2017,Malyzhenkov2020,Marinelli2017,Yan2024}, the peak power has exceeded 1~TW~\cite{Franz2024}, and the pulse repetition rate has increased significantly~\cite{Yan2024}. Moreover, to reduce the size and therefore the cost of XFELs, new methods for accelerating particles and generating electromagnetic radiation have been proposed. In particular, XFELs driven by laser-plasma accelerators are being actively developed~\cite{Nakajima2008,Wang2021,Labat2023,Assmann2020,Nguyen2025}. Futhermore, sources of induced parametric radiation, which arises when electron bunches pass through $\sim100\,\mu$m thick crystals, are proposed as an alternative to undulator FELs~\cite{Baryshevsky1984,Baryshevsky2024}. 

Progress in XFELs is largely due to a deep understanding of particle-field interaction~\cite{Kondratenko1980,Bonifacio1984,Bonifacio1990,Huang2007,Pellegrini2016} and statistical phenomena (shot noise~\cite{Elgin1986,Penman1992,Saldin1998} and velocity spread~\cite{Colson1986}).
To date, software packages are available that take all of these phenomena into account in simulations~\cite{Penman1992,Fawley2002,Reiche1999,Tran1999,Saldin1999,Dejus1999,Freund2000,Campbell2012}.
However, there is still a need for simple models that help us better understand the physical processes occurring in FELs~\cite{Werkhoven1995,Bonifacio1990c,Cai1991,Dattoli2004,Curcio2023,Vinokurov2001,Hemsing2020,Huang2002,Krinsky2004,Maroli1991,Gluckstern1993,Dattoli2007,Piovella1991,Robles2024}. 
Unfortunately, the main attention is paid to models that do not take into account electron velocity spread which, as numerical simulations show, plays an important role in FEL operation~\cite{Robles2024}.

In this regard, the present article is devoted to a simple FEL model that allows for a detailed study the weakly nonlinear particle-field interaction in the presence of electron velocity spread. The paper is organized as follows. In Sec.~II, neglecting radiation slippage and three-dimensional diffraction effects, we will obtain a nonlinear integro-differential equation for the amplitude of the electromagnetic field in a Compton FEL.
In Sec.~III, using numerical solutions of the equation, we will calculate frequency detuning spectra for XFELs driven by laser-plasma accelerators and compare the spectra with those obtained using particle simulations. Sec.~IV contains concluding remarks.

\section{Weakly nonlinear theory}
When neglecting radiation slippage and diffraction effects, the amplification in a Compton FEL is described by the following system of equations~\cite{Bonifacio1990,Hemsing2020}:
\begin{equation}
	\label{eq:theta}
	\frac{d\phi_j}{d\bar z}=\eta_j,
\end{equation}

\begin{equation}
	\label{eq:eta}
	\frac{d\eta_j}{d\bar z}=ae^{i\phi_j}+a^*e^{-i\phi_j},
\end{equation}

\begin{equation}
	\label{eq:a}
	\frac{da}{d\bar z}=-\langle e^{-i\phi_j}\rangle+i\delta a,
\end{equation}
where $a$ is the scaled radiation field, $\phi_j=(k+2\pi/\lambda_u)z_j-ckt$ and $\eta_j=\frac{\gamma_j-\gamma_0}{\rho\gamma_r}$ are the~$j^{\text{th}}$ electron phase and scaled energy, $\bar z=4\pi\rho z/\lambda_u$ denotes the scaled distance along the undulator, and $\delta=\frac{\gamma_0-\gamma_r}{\rho\gamma_r}$ is the frequency detuning. The scaled quantities in \eqref{eq:theta}--\eqref{eq:a} are expressed through the undulator period $\lambda_u$, the speed of light $c$, the electron position~$z_j$ in the undulator, the wave vector~$k$ of electromagnetic field propagating along the system axis, the Pierce parameter $\rho$. In the case of a helical undulator, the quantities $k$ and $\lambda_u$ are related with each other by the expression $k\lambda_u/4\pi=\gamma_r^2/(1+K^2)$. For a planar undulator, the relationship can be written as $k\lambda_u/ 4\pi=\gamma_r^2/(1+K^2/2)$.
In the above formulas, $K$ is a dimensionless parameter proportional to the amplitude of the spatially variable magnetic field of the undulator, while $\gamma_0$ and $\gamma_r\approx\gamma_0$ are the average Lorentz factor of electrons and the resonance Lorentz factor, respectively. 

A few words should be said about equations \eqref{eq:theta}---\eqref{eq:a}. There are several forms of these equations in the literature~\cite{Bonifacio1990,Colson1986,Hemsing2020}. The forms differ in how frequency detuning is taken into account. In some papers, frequency detuning is explicitly present in equation~\eqref{eq:a}, while in others, it is included in the initial values $\eta_{0j}=\eta_j(0)$.

The quantities $\eta_j$ are responsible for the phase change along the undulator axis.
If there is no initial spread of electron velocities, then $\eta_{0j}$ are equal to zero at $\bar z=0$. Otherwise~\cite{Colson1986}, the electron velocities and $\eta_{0j}$ acquire random values $\Delta\beta_{zj}c$ and
\begin{equation}
	\label{eq:eta0j}
\eta_{0j}\approx\frac{k\lambda_u\Delta\beta_{zj}}{4\pi\rho}.
\end{equation}
respectively.
If the beam is monoenergetic, then $\Delta\beta_{zj}\approx-\theta_j^2/2$ is valid~\cite{Colson1986}. (The symbol $\theta_j$ denotes the angle between the undulator axis and the velocity of the $j^{\text{th}}$ particle at the moment of entry into the system.)
If there is no angular spread, then the random quantity $\Delta\beta_{zj}$ can be written as $\Delta\beta_{zj}=(1+K^2)\Delta\gamma_j/\gamma_r^3$ in the case of a helical undulator. For a planar undulator, we can write $\Delta\beta_{ zj}=(1+K^2/2)\Delta\gamma_j/\gamma_r^3$. Here, $\Delta\gamma_j$ is the deviation of the Lorentz factor from the average value $\gamma_0$.

We will assume that, when entering the undulator, there is no field acting on the particle ($a(0)=0$).
Then, the radiative instability begins to grow from small perturbations of the beam density, which are caused by shot noise and (or) e-beam pre-modulation.
The magnitude of these perturbations is determined by the initial value of the bunching factor~$b_0=b(0)=\langle e^{-i\phi_j(0)}\rangle\neq 0$ ($|b_0|\ll1$)~\cite{Hemsing2020}.

Let us try to simplify the system of equations \eqref{eq:theta}---\eqref{eq:a} using reasonable approximations.
As a first step, on the right-hand side of the equation \eqref{eq:eta}, we will neglect the change in $\phi_j$ caused by the interaction of the $j^{\text{th}}$ particle with the radiation field. 
In this approximation, as follows from the equations \eqref{eq:theta} and \eqref{eq:eta}, the phases $\phi_j$ can be expressed through the amplitude $a$ by the double integral
\begin{equation}
	\phi_j(\bar{z})\approx\phi_{0j}+\eta_{0j}\bar{z}+2\text{Re}\int_0^{\bar z}\int_0^{\bar{z}_2}(a(\bar{z}_1)+b_0)\exp(i\phi_{0j}+i\eta_{0j}\bar{z}_1)d\bar{z}_1d\bar{z}_2.
\end{equation}

Substituting the resulting expression for $\phi_j(\bar z)$ into the right-hand side of \eqref{eq:a}, we average the latter over $\phi_{0j}$ and $\eta_{0j}$:
\begin{equation}
	\label{eq:b}
	\langle e^{-i\phi_{j}}\rangle_{\phi_{0j},\eta_{0j}}\approx\int_{-\infty}^{+\infty}\bigg(\frac{1}{2\pi}\int_0^{2\pi}e^{-i\theta_{j}}d\phi_{0j}\bigg)g(\eta_{0j})d\eta_{0j}=\langle iJ_1(2|A_j|)e^{i\arg A_j}\rangle_{\eta_{0j}},
\end{equation}
where $g(\eta_{0j})$ is the distribution function and the following notation is used
\begin{equation}
	\label{eq:Aj}
	A_j=b_0e^{i\eta_{0j}\bar{z}}+e^{-i\eta_{0j} \bar{z}}\int_0^t\int_0^{\bar{z}_{2}}a(\bar{z}_1)e^{i\eta_{0j}\bar{z}_1}d\bar{z}_1d\bar{z}_2.
\end{equation}
Note that separate averaging of $\langle e^{-i\phi_{j}}\rangle_{\phi_{0j},\eta_{0j}}$ over $\phi_{0j}$ and $\eta_{0j}$ means that we neglect correlations between the two quantities.
Thus, we will reduce the entire system of equations \eqref{eq:theta}---\eqref{eq:a} to one integro-differential equation for the field amplitude~\eqref{eq:a}, the right side of which depends in a complex way on $ \eta_{0j}$.

As a next approximation, we replace the right-hand side in \eqref{eq:b} with a function of the mean value $A=\langle A_j\rangle_{\eta_{0j}}$:
\begin{equation}
	\langle J_1(2|A_j|)e^{i\arg A_j}\rangle_{\eta_{0j}}\approx J_1\big(2|A|\big)e^{i\arg A}.
\end{equation}
In the case of small $A_j$, i.e. at the linear stage, this approximation is strictly satisfied, since $J_1(2|A_j|)e^{i\arg A_j}\approx A_j$.

Following~\cite{Colson1986}, we will assume that the electrons have the identical Gaussian distribution over angles in two mutually perpendicular directions orthogonal to the undulator axis. As a result the distribution for $\eta_{0j}$ takes the exponential form:
\begin{equation}
	\label{eq:angularspread}
g_\theta(\eta_{0j})=\frac{e^{\eta_{0j}/\sigma_\theta}}{\sigma_\theta}
\end{equation}
with the mean value
\begin{equation}
	\label{eq:sigmatheta}
\langle\eta_{0j}\rangle=-\sigma_\theta=-\frac{k\lambda_u}{4\pi\rho}\langle\theta_j^2\rangle
\end{equation}
expressed through the angle dispersion $\langle\theta_j^2\rangle$.
In \eqref{eq:angularspread} the random quantity $\eta_{0j}$ accepts only negative values.

In a real beam, in addition to the angular spread, there is also an energy one. As a result, for each group of electrons with the same energy, there should be a shift in the distribution \eqref{eq:angularspread} along the $\eta_{0j}$ axis by a random value $\eta_\gamma$. If energy distribution is gaussian than the distribution of $\eta_\gamma$ is also gaussian~\cite{Colson1986}:
\begin{equation}
	\label{eq:energyspread}
	g_\gamma(\eta_{\gamma})=\frac{e^{-\eta_\gamma^2/2\sigma^2}}{\sqrt{2\pi}\sigma},
\end{equation}
with a standard deviation, as follows from \eqref{eq:eta0j}, equal to
\begin{equation}
	\label{eq:sigma}
\sigma=\frac{\Delta\gamma}{\gamma\rho},
\end{equation}
where $\Delta\gamma/\gamma_0$ is the energy variation.

Averaging the shifted distribution \eqref{eq:angularspread} using \eqref{eq:energyspread}, we find
\begin{equation}
	\label{eq:etaspread}
	g(\eta_0)=\int_{-\infty}^{\eta_0}g_\theta(\eta_0-\eta_\gamma)g_\gamma(\eta_\gamma)d\eta_\gamma=\frac{1}{2\sigma_\theta}e^{(\sigma^2+2\eta_0\sigma_\theta)/2\sigma_\theta^2}\Big(1-\text{Erf}\big((\sigma^2+\eta_0\sigma_\theta)/\sqrt{2}\sigma\sigma_\theta\big)\Big).
\end{equation}
Taking into account \eqref{eq:etaspread} the value $A=\int_{-\infty}^{+\infty}A_jg(\eta_{0j})d\eta_{0j}$ will take the form
\begin{equation}
	\label{eq:A}
	A=\int_0^{\bar z}\int_0^{\bar{z}_{2}}a(\bar{z}_1)\frac{e^{-\sigma^2(\bar{z}-\bar{z}_1)^2/2}}{1-i\sigma_\theta(\bar{z}-\bar{z}_1)}d\bar{z}_1d\bar{z}_2+b_0\frac{e^{-\sigma^2\bar{z}^2/2}}{1-i\sigma_\theta \bar{z}},
\end{equation}
and the integro-differential equation for the dimensionless field amplitude $a$ will be written as follows
\begin{equation}
	\label{eq:ide}
	\frac{da}{d\bar{z}}=i\delta a+iJ_1(2|A|)e^{i\arg A}.
\end{equation}
To solve \eqref{eq:ide}, it is necessary to set four parameters: the initial value of bunching factor $b_0$, the frequency detuning $\delta$, as well as $\sigma$ and $\sigma_\theta$.

Let us note that the linearization \eqref{eq:ide} leads to an equation different from that derived in~\cite{Colson1986}, since the latter deals with different boundary conditions. The integral kernels in~\eqref{eq:A} and~\cite{Colson1986} coincide completely.

\begin{figure}[ht]
	\begin{center}
		\resizebox{80mm}{!}{\includegraphics{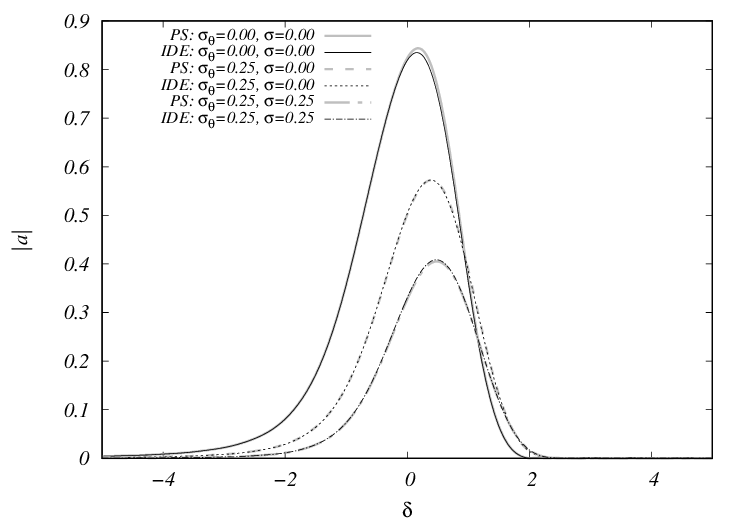}}\\
	\end{center}
	\caption{Frequency detuning spectra for $\bar z=10$.} \label{fig:a10}
\end{figure}
\begin{figure}[ht]
	\begin{center}
		\resizebox{80mm}{!}{\includegraphics{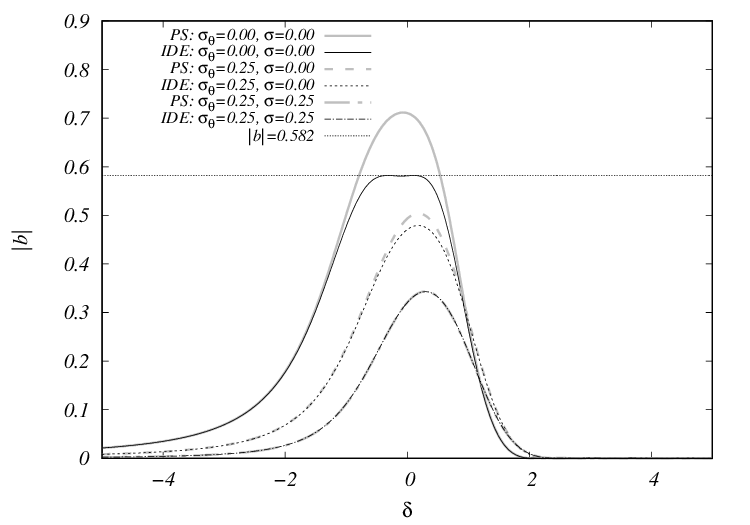}}\\
	\end{center}
	\caption{Dependencies of bunching factor on frequency detuning for~$\bar z=10$.} \label{fig:b10}
\end{figure}
\begin{figure}[ht]
	\begin{center}
		\resizebox{80mm}{!}{\includegraphics{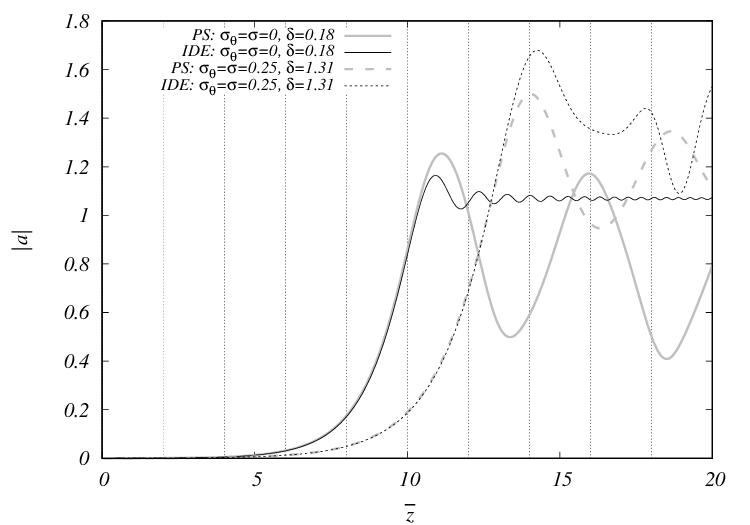}}\\
	\end{center}
	\caption{Field amplification.} \label{fig:az}
\end{figure}
\begin{figure}[ht]
	\begin{center}
		\resizebox{80mm}{!}{\includegraphics{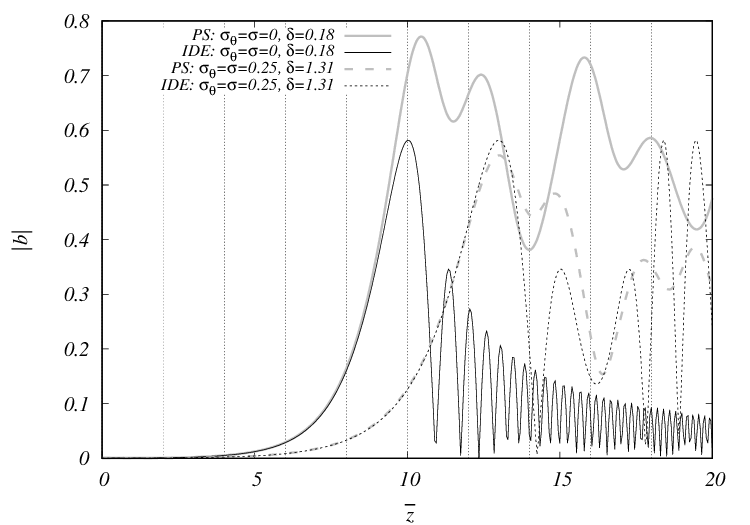}}\\
	\end{center}
	\caption{Bunching process.} \label{fig:bz}
\end{figure}
\begin{figure}[ht]
	\begin{center}
		\resizebox{80mm}{!}{\includegraphics{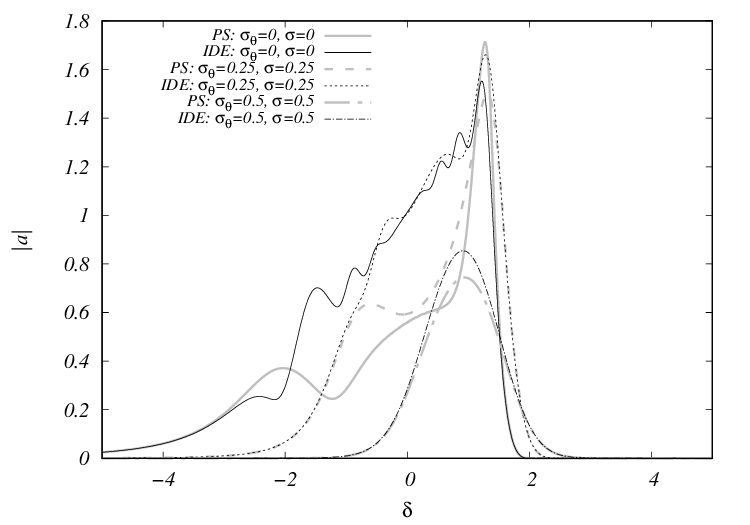}}\\
	\end{center}
	\caption{Frequency detuning spectra for~$\bar z=14$.} \label{fig:a14}
\end{figure}

Let us estimate $\sigma$ and $\sigma_\theta$ typical for XFELs by the example of the AQUA FEL facility with a laser-plasma electron accelerator~\cite{Assmann2020,Nguyen2025}.
The expected values of energy spread and the Pierce parameter are $100\%\cdot\Delta\gamma/\gamma_0\sim0.05$\% and~$\rho\approx1.5\cdot10^{-3}$, respectively, which leads to $\sigma\approx0.3$ (see formula \eqref{eq:sigma}).
The value of $\sigma_\theta$ can be estimated by dividing the square of the normalized beam emittance equal to~$\epsilon_n\approx0.8\cdot10^{-6}$~rad$\cdot$m by $\gamma_0^ 2\sigma_x^2$~\cite{Kim2017}. Here, $\sigma_x$ is the root mean square deviation of one of the transverse electron coordinates.
For the Twiss coefficient $\sigma_x^2\gamma_0/\epsilon_n=5$~m, we obtain $\langle\theta_j^2\rangle\approx6\cdot10^{-11}$rad$^2$ and $\sigma_\theta\approx0. 1$. As a result, we can conclude that typical values of $\sigma$ and $\sigma_\theta$ are tenths of unity.

\section{Detuning spectra}
\label{sec:3}

Using numerical solutions of the integro-differential equation (IDE) derived in Sec. II, we will calculate frequency detuning spectra for different values of $\sigma$ and $\sigma_\theta$. To do this, following~\cite{Hemsing2020}, we choose~$b_0=5\cdot10^{-4}$ as the initial value of the bunching parameter.

Figure~\ref{fig:a10} shows frequency detuning spectra for $\bar z=10$ (in the absence of velocity spread, this case was considered in~\cite{Hemsing2020}).
The choice of $\bar z=10$ is not accidental: at larger~$\bar z$, discrepancies in the bunching factor~$b$ with particle simulations (PS) arise (Figure \ref{fig:b10}). (When solving the integro-differential equation~\eqref{eq:ide}, the maximum value of the bunching parameter is limited by the maximum value of the Bessel function $\text{max}(J_1(x))\approx0.582$.)
At the same time, the frequency detuning spectra calculated using the two approaches are still in good agreement with each other.
This circumstance is primarily due to the short transition time from the linear stage to saturation (Figures~\ref{fig:az} and~\ref{fig:bz}). During the transition, the inaccuracy in~$b$ does not have a noticeable effect on the field amplitude~$a$.

To confirm this, let us compare the frequency detuning spectra obtained by solving~\eqref{eq:ide} with particle simulations for a larger undulator length ($\bar z=14$).
Analysis of the curves shown in Figure~\ref{fig:a14} demonstrates good agreement between the two approaches near the maximums of frequency detuning spectra. As an example, the maximum is achieved at $\delta\approx1.31$ for $\sigma=\sigma_\theta=0.25$.
At this point, the discrepancy between the solution of~\eqref{eq:ide} and particle simulations is about~$10\%$.

In the region of high increments, located near $\delta=0$, the discrepancy between the two theories is significant (Figure~\ref{fig:a14}).
This indicates that the saturation stage has passed and the field amplitude possesses strong oscillations.
The latter can be illustrated by Fig.~\ref{fig:az} if we turn to the curve obtained by the particle simulations at $\delta=0.18$: after passing the point $\bar z\approx11$, the field amplitude oscillates.

Thus, we can talk about good agreement between the weakly nonlinear theory and the particle simulations when analyzing the operation of Compton FELs near saturation. And namely, this case is of the greatest interest from a practical point of view.
It should be taken into account that saturation is achieved for a fixed undulator length at a certain frequency detuning~$\delta$.
If there is a significant deviation from the indicated~$\delta$, one should expect a discrepancy between the weakly nonlinear theory and particle simulations for~$b>0.6$.

\section{Conclusion}
\label{sec:4}
For the field amplitude, an integro-differential equation was derived to describe the operation of a Compton FEL in the presence of electron velocity spread that is typical for modern facilities.
The solutions of the equation denonstrate a good agreement with the results obtained from particle simulations for the bunching factor below~0.6. Moreover, the solutions reproduce the frequency detuning spectrum near its maximum and describe the radiation amplification in the FEL up to saturation.
The agreement with particle simulations is primarily due to the short transition time from the linear stage to saturation. During the transition, the inaccuracy in the bunching factor, which determines the degree of saturation, has little effect on the field amplitude.

The author thanks Professor V.G. Baryshevsky for valuable comments and discussions of the results obtained.

\end{document}